**Electrochemical kinetics of SEI growth on carbon black**

**II: Modeling**

Supratim Das[1], Peter M. Attia[2], William C. Chueh[2] Martin Z. Bazant[1*]

*Corresponding author: bazant@mit.edu

1. Department of Chemical Engineering, Massachusetts Institute of Technology, Cambridge, MA 02139

2. Department of Materials Science and Engineering, Stanford University, Stanford, CA 94305

**Abstract**

Mathematical models of capacity fade can reduce the time and cost of lithium-ion battery development and deployment, and growth of the solid-electrolyte interphase (SEI) is a major source of capacity fade. Experiments in Part I reveal nonlinear voltage dependence and strong charge-discharge asymmetry in SEI growth on carbon black negative electrodes, which is not captured by previous models. Here, we present a theoretical model for the electrochemical kinetics of SEI growth coupled to lithium intercalation, which accurately predicts experimental results with few adjustable parameters. The key hypothesis is that the initial SEI is a mixed ion-electron conductor, and its electronic conductivity varies approximately with the square of the local lithium concentration, consistent with hopping conduction of electrons along percolating networks. By including a lithium-ion concentration dependence for the electronic conductivity in the SEI, the bulk SEI thus modulates the overpotential and exchange current of the electrolyte reduction reaction. As a result, SEI growth is promoted during lithiation but suppressed during delithiation. This new insight establishes the fundamental electrochemistry of SEI growth kinetics. Our model improves upon existing models by introducing the effects of electrochemical

SEI growth and its dependence on potential, current magnitude, and current direction in predicting capacity fade.

## Main

To operate a lithium-ion battery in the thermodynamically stable regime, the electrochemical potential of its electrodes must lie within the stability window of the electrolyte, constraining the open circuit voltage of the full cell. If the negative electrode potential falls below this stability window, a passivating SEI layer will grow on the electrode surface. The SEI layer forms on top of the active negative electrode material, typically graphitic carbon, resulting in depletion of the lithium ion inventory inside the battery[1]. Although the SEI layer is imperative to the stable operation of lithium ion batteries with organic electrolytes in the low potential regime, its continued formation during battery operation is detrimental to all performance metrics of a battery. The growth of this layer presents an additional barrier for ionic and electronic transport within the battery, resulting in resistive heating[2] under high current operations[3] and other unwanted side reactions due to localized thermal effects[4,5]. Furthermore, the increased overpotential for lithium transport can induce lithium plating[6]. Since SEI growth is closely related to the overall degradation of a typical lithium ion battery with a graphitic negative electrode, accurate modeling of this phenomenon is critical for understanding and improving lithium-ion battery lifetime.

***Continuum modeling of SEI growth:*** Continuum modeling of SEI growth involves the interplay of multiple phenomena, any of which could be rate-limiting: (a) electron conduction through the SEI, (b) lithium ion conduction through the SEI, (c) solvent/electrolyte diffusion through pores in the SEI, (d) chemical and mechanical dissolution of the SEI to expose more fresh surface to electrolyte, and (e) charge transfer in SEI growth reaction and its coupling with (de)intercalation and capacitive charge storage. To further complicate this system, the many electrolyte reduction mechanisms result in a variety of products[7] which have different transport properties for ions and electrons. Most modeling work assumes one or two dominant species like LiF ([8]), $Li_2EDC$ ([9]) and $Li_2CO_3$ ([9,10]). This level of detail is usually sufficient to explain the observed trends in capacity fade. However, other factors like electrolyte salt composition and concentration can also influence the composition and performance of the SEI by shifting the equilibrium potential of the formation reactions and influencing the total irreversible capacity during the first few cycles[11,12].

Many models are able to reproduce the experimental 'square-root of time' dependence of capacity fade in the long-term[13–15], but this success does not necessarily translate to the correctness of the rate-limiting step assumed in the SEI growth model. Generally, models of SEI growth assume either electron or solvent transport is rate-limiting; both of these assumptions for the rate-limiting step yield the thickness to be proportional to the square root of time for a system at fixed potential. Early continuum-scale parabolic SEI growth models, such as those by Peled[16] and Broussely et al.[17], assumed electron transport through a homogenous SEI layer is the rate-limiting step. Other authors such as Ploehn et al.[18] have modeled solvent diffusion through the SEI as the rate-limiting step, meaning solvent is consumed at the carbon/SEI interface (i.e. SEI growth takes place at the base). Christensen and Newman[19] considered both lithium and electron transport through the SEI and developed continuum equations in a dilute medium approximation for a single-particle system. Their work was the first to use the notion of a double layer in the SEI/electrolyte interface where the growth reactions were assumed to occur.

More recent modeling work has yet to resolve the controversy over the rate-limiting step. In 2013, Pinson and Bazant[20] developed a single particle growth law of SEI assuming solvent diffusion-limited SEI growing from its base. Notably, the model was integrated into a porous electrode framework to account for spatial inhomogeneities in the lithium ion concentration in the direction of its propagation (i.e. the depth of the electrode). The work demonstrated that substantial variations in SEI thickness as a function of electrode depth arise only under high current operation, and the SEI is otherwise fairly homogeneous in this direction. The base-growth assumption was questioned by Peled and Menkin in a recent review[21], citing physical characterization performed by Edstrom and others[22–26]. These works found that a porous organic layer covers most of the surface of the inorganic, compact SEI, which suggests that solvent molecules do not react at the electrode/SEI interface. An underlying assumption in this argument is that the porous organic layer grows *after* the initial layer solely because it appears to cover the compact layer. This assumption contradicts experimental mass spectrometry observations showing that the porous layer actually grows prior to the underlying non-porous layer[27]. In any case, the question of the location of the growth plane of SEI remains unclear, since the reproduction of the square root of time scaling arises for many standard transport models where the growth rate is limited by the rate at which SEI precursors transport across the SEI.

All of the aforementioned models assume a uniform spatial thickness of the growing passivation layer and a single rate-limiting species. Single particle level inhomogeneities were modeled by Roder et al.[28], using a multi-scale model to capture the spatial non-uniformities of SEI growth. The SEI reaction rates were calculated using atomistic scale simulations of adsorption in a 2D lattice using a kinetic Monte-Carlo (kMC) algorithm described by Burghaus[29] on a selected set of SEI-forming chemical reactions (given in Table 1 of Ref [28]). Single et al.[30] published a coupled electrochemical SEI model that predicts transitions from electron transport limited to solvent-diffusion limited growth after a certain thickness. An explicit porosity variable is used in the model, allowing for a transition in the rate-limiting step as the porosity crosses a critical value. In a subsequent paper[31], the same authors explored the rate-limiting species by studying the limiting long-term growth mechanism (or LTGM). Using experimental data of the potential dependence of SEI growth, they concluded that the diffusion of neutral lithium interstitials, instead of ions or electrons, is the limiting LTGM. This work estimated that the diffusion of neutral lithium occurs at a timescale much larger than the diffusion of lithium ions, since they calculated the ion concentration is higher than that of neutral lithium in SEI interstices[32]. As a result, they concluded that neutral lithium diffusion is important in long-term capacity fade. Thus, different modes of transport (e.g. solvent, ions, electrons or neutral lithium) become important in different timescales for governing SEI growth dynamics.

***Summary of key observations in Part I:*** In Part I of this work, we demonstrated that the post-first cycle growth of the SEI layer shows a strong dependence on voltage, carbon black intercalation current and current direction. Carbon black is used as the model active material for this study due to its high surface area and solid solution behavior during lithium intercalation (see Part I).The SEI layer grows substantially at low potential, consistent with previous work[7,21,25,33], and the irreversible capacity increases with decreasing C rate. While this observation has been attributed to decreasing time per cycle with decreasing C rate[20], the dependence of time-averaged SEI growth on the current (Figure 6b in Part I) shows that cycle time is not the only factor determining SEI growth. We explore the interplay of current and cycle time in our Results and Discussion.

      The most intriguing observation stated in Part I is the strong dependence of irreversible SEI growth on the direction of the carbon intercalation current. SEI growth is considerable

during lithiation of the carbon black electrode, but the growth ceases nearly instantaneously upon reversal of the current. This question of directional asymmetry in intercalation direction at identical potentials is distinct from that of the reversibility of SEI reactions at high potential, which has been previously studied in literature[34–36]. One factor to consider is that the SEI growth process competes for current with the (de)intercalation reaction under constant current conditions, which increases the time per cycle for lithiation and decreases the time per cycle for delithiation. This effect, the time difference in the lithiation and delithiation half-cycles, contributes to the observed 'constant current asymmetry'. However, the stark directional asymmetry in the SEI growth persists even for potentiostatic conditions (Figure 7 of Part I), suggesting another effect is at play. This observation is not captured by the commonly accepted models of electrochemical SEI growth published to date.

In Part II, we investigate the hypotheses proposed in Part I to explain the observed directional asymmetry. We mainly explore the third hypothesis, which treats the SEI as a mixed ion-electron conductor (MIEC). In this model, the exchange current and overpotential of the electrolyte reduction reaction are functions of the lithium ion and electron concentrations in the SEI. Specifically, the electronic conductivity of the SEI is a dynamic parameter dependent on both current rate and direction through its coupling to the lithium-ion concentration. We explore different dependencies of the electron conductivity $\tilde{\sigma}_{e^-,SEI}$ on the absorbed Li$^+$ ion concentration $\tilde{c}_{Li^+,SEI}$ to understand the electron-ion coupling that drives the kinetics of SEI growth. Parametric analyses are used to study the rate-limiting step for outer SEI growth, which highlights the importance of the electron conduction mechanism in the SEI. This work presents a new framework for modeling the SEI that captures the experimental observations of Part I, specifically the directional asymmetry.

                                                        *Das et al.*

## Model Description

The SEI layer growth process in the post-first-cycle regime depends upon numerous coupled electrode and particle scale phenomena, all of which depend on the potential and the concentrations of lithium ions, solvent molecules, and electrons at a particular point in the domain. A schematic of the dominant factors influencing post-first-cycle SEI growth on a carbon electrode is shown in Figure 1. The one-dimensional SEI domain originates at the carbon black/SEI interface and increases rightward in the schematic. The base of the SEI represents $x = 0$ and the SEI/electrolyte interface represents the $x = \tilde{L}$ boundary, where $\tilde{L}$ is the dimensionless SEI thickness at any point in time.

Some key features of this model framework are as follows:

i.   This work endeavors to understand the electrochemistry of SEI growth in a small, well-defined time scale: the second cycle to the first few cycles, dubbed 'post-first-cycle' regime, an important regime in lithium ion battery degradation. The SEI formation process on the first cycle is distinct and not considered in this study.

ii.   The model is agnostic to the details of individual reactions leading to SEI growth; the equilibrium potential of SEI growth is obtained using data from experimental measurements. This "effective" equilibrium potential is a lumped value that incorporates the multitude of solvent/electrolyte reduction reactions, each with their own distinct equilibrium potentials[8–10].

iii.  Experimental work reported in Part I of this communication uses an electroanalytical technique to isolate the electrochemical signature of SEI growth on carbon black during galvanostatic cycling using $\varDelta dQ/dV$ as the measurement output. We compare our model results to the experimentally-measured values of $\varDelta dQ/dV$ and use this comparison as the error metric to evaluate model performance, instead of an integrative property like capacity.

iv.  All electrochemical reactions described in the model consider the effect of changing species concentrations (specifically, electrons and ions) at the interfaces in which they occur. This allows us to study the effect of ion and electron concentrations on fluxes at the reaction interfaces.

***Model formulation:*** Since the model is motivated by the key experimental results discussed in Part I, we make use of these insights to simplify the picture. The diffusion and transport of lithium ions through the porous and non-porous SEI layers (process 2b in Figure 1) is generally accepted to be a fast process during early cycles, as evidenced by the low interfacial impedance attributed to the SEI layer. Proposed mechanisms of lithium ion transport through the SEI include transport through interstices and grain boundaries by a 'knock-off' mechanism[32,37,38]. Based on the experimental observation of strong voltage dependence for SEI growth, we assume that electron transport is rate limiting as opposed to solvent diffusion (process 3 a-b in Figure 1) for 'post-first-cycle' early stage SEI, since the former is expected to exhibit stronger voltage dependence.

In this picture, the (de)intercalation reaction takes place at the electrode/SEI interface where an electron supplied from the electrode reacts with a lithium ion diffusing through the SEI from the electrolyte. Electrons also move across the SEI layer to reach the outer electrolyte interface, where they react with lithium ions and electrolyte (solvent and/or salt) to form more SEI as a parasitic reaction that passivates the surface. We model the SEI-forming Faradaic reaction using the classical Butler-Volmer equation,

$$I_{SEI} = I_{SEI}^0 \left[ e^{(1-\alpha_{SEI})\tilde{\eta}_1} - e^{-\alpha_{SEI}\tilde{\eta}_1} \right] \tag{1}$$

where the (dimensionless) overpotential $\tilde{\eta}_1 = \Delta\tilde{\phi}_S - \Delta\tilde{\phi}_s^{eq}$ is the potential drop $\Delta\tilde{\phi}_s$ across the compact Stern layer at the SEI/electrolyte interface, relative to its equilibrium value $\Delta\tilde{\phi}_s^{eq}$ given by the Nernst equation. We set the charge-transfer coefficient $\alpha_{SEI}$ to 0.5, consistent with most models and Marcus theory for outer-sphere electron transfer[39], although we explore the possibility of Butler-Volmer kinetics that is asymmetric between cathodic and anodic SEI growth/stripping reactions (not be confused with asymmetry upon carbon lithiation and delithiation).

We neglect the diffuse part of the electric double layer in the electrolyte by taking the Helmholtz limit[40,41], where the Debye screening length is much smaller than the effective Stern layer thickness, which is reasonable for concentrated battery electrolytes. In this limit, the Stern layer voltage drop that drives the SEI formation reaction is given by $\Delta\tilde{\phi}_S = \tilde{\phi}_{elec} - \tilde{\phi}_{SEI}(x = \tilde{L})$, where $\tilde{\phi}_{SEI}(x = \tilde{L})$ is the potential on the outer surface of the SEI layer and $\tilde{\phi}_{elec}$ is that of the electrolyte just outside the double layer, which is assumed constant in this study due to the low applied currents (<C/5). More generally, our equations could be coupled with electrode and cell models that includes concentration polarization in the electrolyte to account for large applied currents[42,43] following Pinson and Bazant[20].

A critical aspect of our model is to view the SEI layer as a MIEC with time-varying concentrations of both mobile electrons $\tilde{c}_{e^-,SEI}$ and inserted ions $\tilde{c}_{Li^+,SEI}$. Accounting for all species involved in the SEI growth reaction, the Nernst equation for the equilibrium Stern layer voltage drop then reads,

$$\Delta\tilde{\phi}_s^{eq} = \Delta\tilde{\phi}_S^0 + \ln\left(\frac{\tilde{c}_{Li^+,SEI}\tilde{c}_{e^-,SEI}}{\tilde{c}_{SEI}}\right) \tag{2}$$

where the reference value $\Delta\tilde{\phi}_S^0$ is related below to the electrode potential for the onset of the SEI reaction relative to lithium metal reference (see Table 1). In principle, the dimensionless SEI concentration (a measure of SEI product fraction per unit thickness) $\tilde{c}_{SEI}$ could vary as the composition or porosity of the layer changes, but here we set $\tilde{c}_{SEI} = 1$ as a first approximation. In Equation (2), all activities have been replaced by dimensionless concentrations (for a dilute solution within the SEI), scaled to a common reference value consistent with the definition of $\Delta\tilde{\phi}_S^0$. In the same spirit, we must also account for both ion and electron concentrations in the derivation of the Butler-Volmer equation (1), leading to a consistent definition of the exchange current density for the SEI reaction that also includes the local concentration of mobile electrons[44],

$$I_{SEI}^0 = k_{SEI}\left(\tilde{c}_{e-,SEI}\,\tilde{c}_{Li^+,SEI}\right)^{(1-\alpha_{SEI})}\tilde{c}_{SEI}^{\alpha_{SEI}} \tag{3}$$

where $k_{SEI}$ is a dimensionless rate constant. Although we neglect the dependence of the SEI growth reaction on the specific composition of the SEI matrix, we do account for its dependence on the concentrations of mobile electrons and inserted lithium ions, which are strongly coupled to carbon black intercalation and SEI growth kinetics.

We recognize that space charge may play a role, resulting from imbalances in the concentrations of electrons, ions, and fixed charges in the SEI matrix, which are governed additionally by Poisson's equation. Here, as a first approximation, we assume local electroneutrality. Moreover, we also assume intrinsic defect chemistry. Together, these assumptions result in equal electron and ion concentrations, i.e. $\tilde{c}_{Li^+,SEI} = \tilde{c}_{e-,SEI}$, everywhere within the SEI.

Although we do not explicitly model charge as a first approximation, our model can be derived as the self-consistent thin double-layer limit of more general "Frumkin-Butler-Volmer" models of electrochemical kinetics coupled with diffuse interfacial charge[41,45–48]. Even with our level of description, the implied charge stored capacitively across each layer in the model could be calculated *a posteriori* by multiplying each voltage drop by the capacitance of that layer. Assuming all charge resides in thin interfaces between charge-free dielectric layers, the capacitance of each layer could be estimated as its dielectric constant divided by thickness. In this way, the implied surface charge density at the SEI/electrolyte interface could be calculated from the Stern layer voltage drop as $q_s = \frac{\epsilon_S \Delta \tilde{\phi}_S}{h_S}$ , where $\epsilon_S$ is the dielectric constant and $h_s$ the thickness of the Stern layer, although this is not required for the SEI growth model. Finally, although the mobile ions and electrons are strongly coupled by Coulomb forces and exhibit ambipolar diffusion while maintaining electroneutrality[49], we neglect the associated concentration polarization by assuming that the currents are far below the diffusion-limited current in the SEI layer.

Since electrons do not cross into the electrolyte, the Faradaic reaction current for SEI growth must be equal to the electronic "leakage current" across the SEI layer. Consistent with the assumptions above, the latter is given by Ohm's law,

$$I_{SEI} = \frac{\Delta \phi_{SEI}}{R_{SEI,e}} \tag{4}$$

where $R_{SEI,e}$ is the electronic resistance of the SEI per unit area of active electrode material and $\Delta \phi_{SEI} = \phi_{SEI}(x = \tilde{L}) - \phi_{SEI}(x = 0)$ is the voltage drop across the SEI layer. To capture trends over many cycles, the SEI resistance may depend nonlinearly on the dimensionless thickness $\tilde{L}$ as a power law to indirectly account for changing porosity or fractal dimension,

$$R_{SEI,e} = R_{SEI,e}^0 \frac{\tilde{L}^\beta}{\tilde{\sigma}_{SEI,e}(\tilde{c}_{Li^+,SEI})} \tag{5}$$

where $R_{SEI,e}^0$ is a reference electronic resistance for the initial SEI layer ($assumed\ \tilde{L} = 1$) and $\tilde{\sigma}_{SEI,e}$ is the dimensionless mean electronic conductivity, assumed to be approximately uniform across the SEI layer but varying in time with the mobile electron concentration $\tilde{c}_{e^-,SEI}$, or the ion concentration $\tilde{c}_{Li^+,SEI}$ by electroneutrality.

***Electronic conductivity:*** The SEI electronic conductivity depends on the mobile electron concentration, which is equal to the inserted ion concentration by the assumption of electroneutrality in the intrinsic limit discussed above. For simplicity, we postulate that the electronic conductivity exhibits a power-law dependence on the lithium ion concentration:

$$\tilde{\sigma}_{SEI,e}(\tilde{c}_{Li^+,SEI}) = \tilde{c}_{Li^+,SEI}{}^{\nu} \tag{6}$$

The exponent $\nu$ allows us to explore different modes of electron conduction, including three limiting cases from semiconductor physics:

1) <u>Constant conductivity</u> ($\nu = 0$): In this case, the SEI layer has a constant electronic conductivity irrespective of lithium-ion concentration and potential. Physically, this corresponds to the situation where the SEI cannot insert lithium ions, or that the electron concentration is fixed by impurities.

2) <u>Ideal mixed ion-electron conductor</u> ($\nu = 1$): In this case, each inserted ion adjusts the Fermi level so as to enable another mobile electron to enter the material and carry electronic current, in proportion to the local ion concentration. The SEI electronic conductivity is thus linear with lithium-ion concentration. Physically, this corresponds to a semi-metal, where the additional mobile electrons associated with ion insertion are sufficiently delocalized and concentrated that each contributes equally to the total conductivity.

 *Das et al.*

3) <u>Non-ideal mixed ion-electron conductor</u> $(\nu = 2)$: In this case, the electronic conductivity increases nonlinearly with inserted ion concentration, as in an semiconductor with hopping conduction of electrons across percolating clusters of localized states[50–52]. As the concentration of ions is increased, a long-range conductive pathway throughout the material is established beyond a certain critical fraction[53] of occupied sites, where the largest cluster size becomes comparable to the thickness[65]. For an infinite system near the critical point, the effective conductivity $\sigma$ as a function of the inserted ion fraction $p$ scales as, $\sigma \propto (p - p_c)^t$, where $p_c$ is the critical threshold filling fraction. The scaling exponent $t$ is universal (depending only on the embedding dimension) and close to 2 in three dimensions[50,54,55]. For small, finite systems, such as the early SEI layer, the percolation transition is smoothed, and the conductivity may be approximated by a simple quadratic function across the full concentration range, as shown in Figure 2.

In the absence of any direct measurement or predictive simulations of electronic conductivity, we fit the prefactor in each case to the experimental $\Delta dQ/dV$ data as a function of voltage (and indirectly to state of charge). By combining Equations (1) - (6), the voltage drops across the SEI and Stern layers are connected by the conservation of electronic current, which leaks through the SEI layer to cause electrolyte reduction that further grows the SEI, in a way that depends on the concentration of inserted lithium ions.

***Lithium ion insertion and transport in the SEI and carbon electrode:*** To complete the model, we postulate mechanisms for lithium ion insertion into and transport through the SEI layer, followed by Faradaic reaction and intercalation into the active carbon material. To determine the lithium concentration in SEI, we adopt the standard model of surface charge regulation by quasi-equilibrium ion absorption from the electrolyte, which has been successful in diverse applications[47,56–60]. By equating the electrochemical potential of ions in the SEI layer and the electrolyte, assuming each is a dilute solution, we arrive at the absorption isotherm,

$$\tilde{c}_{Li^+,SEI} = e^{-\Delta\tilde{\phi}_S} \; e^{-E_{ads}/(k_BT)} \tag{7}$$

where $E_{ads}$ is the specific energy of absorption of lithium ions from the electrolyte into the SEI. The modulation of the ion concentration in SEI, and thus its electronic conductivity, by the Stern layer voltage is an important source of asymmetry in the model.

Inserted lithium ions migrate through the SEI from the electrolyte until they reach the SEI/electrode interface, where they participate in the dominant, reversible Faradaic reduction reaction. Although our SEI model could be applied to any electrode reduction, such as lithium metal electrodeposition instead of lithium ion intercalation, we focus here on the experiments of Part I for Li-ion insertion in carbon black as a model electrode for SEI growth in Li-ion batteries. The first step in modeling the electrode is to describe the open circuit voltage via the activity $a_R$ of the reduced state, or, in this case, the electrochemical potential of intercalated lithium ions $\mu_{int,Li^+}(\tilde{c}) = k_B T \ln a_R(\tilde{c})$, which depends on their concentration $\tilde{c}$ according to an appropriate thermodynamic model of the solid[44]. Neglecting any variable polarization of the counter-electrode, the open-circuit potential can be modeled using the Nernst equation for the half reaction of one-electron reduction, $O + e^- \leftrightarrow R$, as,

$$\Delta\tilde{\phi}_{int,eq} = \Delta\tilde{\phi}_{int}^0 - \ln\left(\frac{a_R}{a_O}\right) = \Delta\tilde{\phi}_{int}^0 - \mu_{int,Li^+}(\tilde{c}) - \ln\tilde{c}_{Li^+,SEI} \tag{8}$$

where $\Delta\tilde{\phi}_{int}^0$ is the standard redox potential for intercalation, relative to the ideal counter-electrode (e.g. lithium metal), and $a_O$ is the activity of the oxidized state of the reaction (the lithium ion in SEI), which we set equal to the dimensionless ion concentration, consistent with the dilute solution approximation above.

                                                    *Das et al.*

Lithiated carbon black exhibits solid solution behavior and a suppression of graphitic phase separation[61], as confirmed by in-situ X-ray diffraction[62]. Therefore, we assume a semi-empirical chemical potential of the form,

$$\mu_{int,Li^+}(\tilde{c}) = A\,ln\left(\frac{\tilde{c}}{1-\tilde{c}}\right) - B\tilde{c}^C \tag{9}$$

where the first term is the (dimensionless) configurational entropy of an ideal solid solution of intercalated ions and vacancies and the second term is an empirical approximation for the dimensionless enthalpy of intercalation. Fitted values of the parameters $A$, $B$ and $C$ are given in Table 1 for the open circuit potential of lithium in carbon black relative to lithium metal.

The Faradaic intercalation reaction is again described using Butler-Volmer kinetics,

$$I_{int} = I_{int}^0 \left[ e^{(1-\alpha_{int})\tilde{\eta}_2} - e^{(-\alpha_{int})\tilde{\eta}_2} \right] \tag{10}$$

where the intercalation overpotential $\tilde{\eta}_2$ is given by $\tilde{\eta}_2 = \Delta\tilde{\phi}_{int} - \Delta\tilde{\phi}_{int,eq}$ and the intercalation exchange current is given by,

$$I_{int}^0 = k_{int}\left(\tilde{c}_{SEI,Li^+}\right)^{(1-\alpha_{int})}\tilde{c}^{\alpha_{int}} \tag{11}$$

where $k_{int}$ is another rate constant, and $\alpha_{int}$ captures the asymmetry between carbon lithiation and delithiation reactions. The carbon black intercalation kinetics is assumed to be reaction limited.

The model is completed by the conservation of ionic current. We operate our simulations in galvanostatic mode. The sum of the intercalation and SEI electronic currents is constrained to be equal to the total driving current in the cell as,

$$I_{tot} = I_{int} + I_{SEI} \tag{12}$$

While the SEI growth "leakage" current $I_{SEI}$ is carried by electrons across the SEI layer, the intercalation current is carried by the ions,

$$I_{int} = \frac{\Delta\phi_{SEI}}{R_{SEI,Li+}} = \frac{\Delta\phi_{SEI}\tilde{c}_{Li^+,SEI}}{R^0_{SEI,Li+}} \tag{13}$$

which encounter their own ohmic resistance across the SEI layer, $R_{SEI,Li+} = R^0_{SEI,Li+}/\tilde{c}_{SEI,Li+}$ in parallel with that of the electrons. For typical situations where $I_{int} \gg I_{SEI}$, we expect the SEI layer to be much more resistive to electrons than to ions, $R^0_{SEI,e} \gg R_{SEI,Li+}$, although these are left as fitting parameters. Since ions encounter much less resistance than electrons, the transport mechanism is not expected to affect the overall simulation. Thus, we take ionic conductivity in the SEI to be linearly proportional to the carrier concentration, $\tilde{c}_{SEI,Li^+}$. Finally, we consider the mass balance of SEI growth driven by the electronic leakage current,

$$\frac{d\tilde{L}}{dt} = \frac{I_{SEI}}{e} A_{SEI} \tag{14}$$

where $A_{SEI}$ is the specific surface area of the typical SEI reaction product in contact with the electrolyte. In our simulation, it is assumed that SEI grows uniformly on the active material, and thus $A_{SEI}$ is assumed equal to the specific surface area of active material. Values of all tunable parameters used in this model are tabulated in Table 1.

***Numerical method and validation:*** We solve all the equations on a static equidistant one-dimensional grid within the SEI layer. The $x = 0$ of the grid is located at the base of the SEI layer on the carbon black/SEI interface, while $x = \tilde{L}$ is the moving outer boundary of the SEI (the SEI/electrolyte interface). To solve this system of equations, we take the general approach of discretizing each in space using a finite difference method to obtain a system of differential algebraic equations (DAEs), and then stepping in time using a variable-order adaptive time stepper (MATLAB's ode15s function). All equations are solved for the six primary variables - $\Delta\tilde{\phi}_{int}$, $\Delta\tilde{\phi}_{SEI}$, $\Delta\tilde{\phi}_{S}$, $\tilde{c}_{Li^+,SEI}$, $\tilde{\sigma}_{SEI,e}$ and $\tilde{L}$ - simultaneously in the whole domain at all times. The stopping criteria on lithiation and delithiation are the upper and lower cutoff potentials of 1.2 V and 0.01 V, respectively; the upper cutoff potential is taken to be 1.2 V for model analyses since all SEI growth occurs at potentials below this value (see Figure 4 in Part I). The parameters in Table 1 are obtained from differential capacity ($dQ/dV$) data of single cells for lithiation and delithiation across a variety of galvanostatic current values ranging from C/100 to C/5. The goodness of fit for each iteration is determined using partial least squares, and MATLAB's fmincon function used to obtain the optimal values.

**Summary of conventions:** In this work, reduction (or intercalating) current is assumed to be negative, consistent with the IUPAC convention. Negative current corresponds to discharging of the carbon black/lithium half cell. All concentration terms in this model are non-dimensionalized using the electrolyte bulk lithium ion concentration $c_0$ (assumed 1 M in our electrolyte, 1.0 M LiPF$_6$ in 1:1 wt% EC:DEC), and potentials with the thermal voltage, $k_B T/e$. All battery operations are considered isothermal, operating at T = 298K. All currents and exchange current prefactors are defined per unit area of active material. All potentials are defined with respect to the standard potential of lithium metal as the reference.

**Summary of model framework:** In this model, intercalation is coupled to SEI growth via the species concentrations which appear in the overpotentials ($\tilde{\eta}_1, \tilde{\eta}_2$) and the exchange currents ($I_{int}^0, I_{SEI}^0$) and via the constraints in Equations (4) and (13). Importantly, this coupling exists even for thin SEI where the ionic resistance is negligible. The direction of the current determines the sign of the polarization of the SEI and outer film, thereby controlling the lithium ion concentration in the SEI via Equation (7). This concentration term affects the exchange currents and overpotentials of both reactions and makes the reaction kinetics sensitive to current direction and magnitude. The resistance of the SEI to electrons ($R_{SEI,e}$), a measure of the degree of passivation, is dependent on the potential, current magnitude, and current direction, as the mechanism of electron transfer depends upon those factors. Since the SEI electronic resistance determines the electron flux to the outer film that participates in electrolyte decomposition, it effectively controls the available overpotential of the SEI growth reaction.

***Summary of assumptions/simplifications:*** This modeling endeavor is inspired by the experimental findings outlined in Part I and undertakes an approach that uses only the level of detail that can be supported directly using the experimental data, minimizing the number of parameters. We briefly summarize some key assumptions here:

*Outer surface growth of SEI:* One of the biggest challenges in SEI growth modeling is obtaining conclusive evidence of the location of its growth plane. Motivated by the strong voltage dependence observed experimentally, we assume that the SEI layer grows from its outer surface (i.e. electron-transport limited) and formulate our rate-limiting step hypotheses accordingly. This allows us to effectively model the potential dependence and coupling between intercalation and SEI growth.

*Uniform SEI on single particle:* The model is area-averaged (i.e. 1-dimensional) and assumes SEI of uniform thickness to be growing on top of a single particle of carbon black, which is a simplification of the actual scenario of SEI growth in porous electrodes.

*Intercalative charge storage in carbon black:* Carbon nanomaterials such as carbon black typically have high specific capacitance due to the high specific surface area[63]. While both intercalation and capacitance contribute significantly to charge storage in carbon black (elucidated in Part I), for simplicity we assume the carbon black capacity is purely intercalative.

*Electrolyte sourced lithium ions:* In this work, we assume the only source of lithium ions participating in SEI growth is the electrolyte, as opposed to the intercalated lithium in the lithiated electrode. The latter source can be expected to become important under long periods of storage, where chemical SEI growth reactions may occur under open circuit conditions (affecting "calendar life").

*Butler-Volmer kinetics:* As in most electrochemical engineering models, we use the empirical Butler-Volmer equation to model the faradaic reaction kinetics of intercalation and SEI formation, which can be justified at low overpotentials but may over-estimate the reaction rate at high overpotentials, compared to quantum mechanical models of electron transfer, such as Marcus kinetics[64,65].

*Uniform lithium ion concentration profile in SEI:* We assume a spatially uniform concentration profile of lithium ions in the SEI layer in this time regime, where the thickness of the layer is small. Effects due to diffusion driven concentration gradients usually occur over longer timescales than that of the system under study. Therefore, this assumption is unlikely to produce significant errors in the model predictions. However, the errors in prediction may increase at larger SEI thicknesses or evolution of sufficiently porous SEI[30].

*Moderately dilute solution approximation:* While we do account for concentration dependencies of the exchange current and the overpotential for both intercalation and SEI formation, we set the activity coefficients to unity for all species, as in a dilute solution. On the other hand, we assume the solution is concentrated enough (as in typical battery electrolytes) to neglect diffuse double layers and associated Frumkin corrections[40,45] to Butler Volmer kinetics.

## Results and Discussion

In Part I, we isolate the contribution of the SEI to the differential capacity, $dQ/dV$, by subtracting the differential capacity in an early cycle from the differential capacity at a later "baseline" cycle. The differential capacity in an early cycle (e.g. cycle 2) has contributions from both carbon black and SEI, while the differential capacity in the baseline cycle primarily measures carbon black. Thus, the resultant difference, $\Delta dQ/dV$, measures the voltage dependence of SEI growth within the early cycle. We use $\Delta dQ/dV$ as the experimental quantity to evaluate goodness of fits with model predictions. In this section, we evaluate the second and third hypotheses described in Part I to understand the nature of the rate-limiting step of post-first-cycle SEI growth. Briefly, the second hypothesis states that directional asymmetry of SEI growth on carbon lithiation vs. delithiation could arise from the intrinsic differences between lithiation and delithiation into carbon, phenomenologically captured through deviations of $\alpha_{int}$ from 0.5 in our model. The third hypothesis states that the SEI behaves as a MIEC, and a concentration-dependent electron conductivity is the cause of directional asymmetry in SEI growth. The outcome guides our understanding of the rate-limiting process that governs the kinetics of SEI growth.

***Constant current asymmetry:*** Figure 3 shows the current and dimensionless SEI thickness as a function of time for the case of the SEI behaving as a simple resistor with constant electronic conductivity, i.e. the exponent $\nu = 0$ in Equation (6). From Figure 3b, post-first cycle SEI clearly does not follow a $t^{\frac{1}{2}}$ growth, as is widely observed for long term SEI growth. SEI growth is low in the high potential region (i.e. beginning of lithiation and end of delithiation in Figure 3) and rapidly accelerates as the potential decreases below $\lessgtr$ 0.3 V. The constant conductivity SEI model, though not in quantitative agreement with experiments, does demonstrate some degree of directional asymmetry – in this case, the growth in the delithiation step is 27% of that in the lithiation step. During lithiation, as the SEI current reduces the total current available for intercalation (Figure 3a) under a constant current constraint (Equation (12)), the time required to reach the lower cutoff potential of 10 mV is longer than what would be expected for a nominal driving current of C/10 (i.e. 10 hours). Upon switching of the current direction, the deintercalation current becomes larger in magnitude than the total current because the SEI current remains negative at the switching potential of 10 mV. Higher deintercalation current leads to the system reaching the upper voltage cutoff of 1.2 V more quickly than the expected 10 hours for a C/10 nominal driving current. Less delithiation cycle time leads to reduced growth of SEI, and this effect can be thought of as 'constant current asymmetry'. A large SEI current, which is directly proportional to the specific surface area of active material, results in a more pronounced 'constant current asymmetry'. For active materials with micron-scale particle sizes like graphite, this effect will be less significant compared to nanostructured negative electrode materials. This effect causes slower charging and faster discharging during galvanostatic cycling for any system with non-negligible SEI growth.

***Determination of rate-limiting mechanism of post-first-cycle SEI growth:*** However, we observed experimentally that the directional asymmetry of SEI growth persists under potentiostatic conditions as well (see Figure 7 in Part I). Thus, this directional asymmetry has a deeper physical origin than the trivial coupling discussed in the previous section. To quantitatively capture the directional asymmetry, we explore other modes electronic conduction in SEI (i.e., $\nu = 1$ and 2 given in Equation (6)). The colormap plots in Figure 4 show the degree of asymmetry, defined as difference between the lithiation and delithiation SEI capacities ($Q_{diff}$), as a function of four parameters: SEI resistance to electrons $R_{SEI,e}$, the SEI reaction prefactor $k_{SEI}$, the correlation exponent $\nu$ and the transfer coefficient $\alpha_{int}$ of (de)intercalation. The upper bound of $k_{SEI}$ is the value at which the SEI current approaches the total current. The SEI capacities $Q_{lith}$ and $Q_{delith}$ are calculated by integrating the 2nd-cycle $\Delta dQ/dV$ with respect to voltage from 0.01 V to 0.7 V. The shaded regions describe the range of parameter values that reproduce the $Q_{diff}$ values for experimentally observed average $Q_{lith}$ and $Q_{delith}$ values given in Figure 7. Some features common to all three columns are worth pointing out as they lead to important insights:

First, $Q_{diff}$ is very sensitive to $R_{SEI,e}$ since a more insulating SEI results in a higher degree of 'constant current asymmetry'. As described earlier, constant current asymmetry increases the time spent in lithiation and reduces the time spent in delithiation.

Second, the gradient of the change in color is almost always parallel to the SEI current prefactor ($k_{SEI}$) axis for the range of explored values, indicating that $Q_{diff}$ is a weak function of $k_{SEI}$. From the Butler-Volmer equation, the SEI current is identical for a given overpotential, irrespective of the current direction. We note that the fitted SEI equilibrium potential is 0.73V, which means that we are always on the cathodic branch of the Butler-Volmer equation.

Third, the impact of the carbon (de)intercalation charge-transfer coefficient, $\alpha_{int}$, on directional asymmetry is fairly small. A deviation of $\alpha_{int}$ from 0.5 physically could correspond to a difference in the free energy landscape of intercalation and de-intercalation pathways (the second hypothesis given in Part I), or in the context of asymmetric Marcus-Hush kinetics of electron transfer, a difference in the solvent reorganization free-energy curvatures for the reduced and oxidized states[66-68]. However, under the assumptions given in this paper, this hypothesis cannot quantitatively explain the experimental observations.

The colormap plots in column 1 of Figure 4 depict that the constant electronic conductivity ($\nu = 0$) model displays very low directional asymmetry across a range of different parameters, which is inconsistent with experimental observations. Thus other modes of electron conduction need to be explored. Column 2 of Figure 4 shows a parametric study of directional asymmetry using the ideal MIEC model ($\nu = 1$). This model manages to partially reproduce the observed directional asymmetry, though for a narrow range of parameters in the case of $\alpha_{int} = 0.7$ for nominal C rates of C/100 and C/50. Importantly, these 'agreements' (refer to Figure 7 for experimental data of SEI capacities) are only obtained when using different parameters for different C rates in this parametric analysis, and thus cannot be used to develop a consistent theory explaining directional asymmetry in SEI. In addition, the microscopic Marcus theory[66] predicts a symmetric charge transfer coefficient $\alpha_{int}$ equal to 0.5 for a Butler-Volmer formulation[69-71] of ion intercalation kinetics for moderate overpotentials of less than 1 V ($<\lambda$, the reorganization energy)[72]. A value of $\alpha_{int}$ not equal to 0.5 (such as $\alpha_{int} = 0.7$ shown in Figure 4, which physically implies a greater bias for reduction current) may often mask some diode-like physics related to space charge at the interface[44,64-66,68], which causes the intercalation current, and consequently the electrolyte/solvent reduction current, to be higher for one polarization over the reverse. We suspect may be the case in our system since the $Q_{diff}$ for $\alpha_{int} = 0.7$ matches with experimental values for a narrow range of $R_{SEI,e}$ for the $\nu = 1$ case (see middle subplot of bottom row in Figure 4).

We account for this diode-like behavior explicitly by considering a non-ideal MIEC model ($\nu = 2$), and set $\alpha_{int}$ equal to 0.5. A parabolic dependence physically corresponds to the effect of percolation for hopping conduction of electrons across the SEI, as described earlier. The asymmetric nature of the functional form of Equation (7) describing the absorbed lithium ion concentration implies that the concentration is enhanced in one direction of the polarization. This effect results in an electron conductivity that is directional in nature. Column 3 of Figure 4 shows the colormap plots for the percolation-based non-ideal mixed ion-electron conduction effect. In this case, we obtain quantitative agreement (within error) using a symmetric (de)intercalation charge transfer coefficient $\alpha_{int} = 0.5$ for a value of $R_{SEI,e} \sim 10 - 30\ \Omega$. The best fit with experiments across all nominal C rates is obtained for a reference electron conductivity of $3.5 \times 10^{-11}$ S m$^{-1}$ which gives an SEI resistance $R_{SEI,e} = 14\ \Omega$ for a measured specific surface area of $\sim 62$ m$^2$ g$^{-1}$ for the carbon black electrode. Figure 5 demonstrates the good agreement of the theoretical $\Delta dQ/dV$ predictions with experiments for a range of currents from C/100 to C/5. The simulation results in Figures 5 and 6 are generated using the parameters mentioned in Table 1 and assuming a charge transfer coefficient of 0.5 for both carbon (de)intercalation and SEI electrolyte reduction reactions. The strong goodness-of-fits suggest that the inclusion of this nonlinear dependence of electron conductivity on the lithium ion concentration is the most important factor in our model determining the rate of post-first-cycle outer SEI growth for moderate driving currents. This result adds support to our main hypothesis that the post-first cycle SEI growth is limited by the flux of electrons across the layer, which affects the electron availability in the outer reaction film.

**Mechanism of electron conduction across the SEI layer:** The SEI, often considered an electronic insulator, may behave like a non-ideal MIEC if the concentration of inserted ions increases beyond a certain threshold to facilitate hopping conduction among the localized nearby electronic states[73]. From Figure S1 in the Supplementary Material, we find that the Tafel slope is best reproduced for the quadratic dependence of conductivity on the lithium concentration. We draw from this and the results obtained in Figures 4 and 5, to develop a theory that considers how both an asymmetry in the lithium ion concentration in the SEI and the nonlinear dependence of electron conductivity on the lithium ion concentration lead to the observed directional asymmetry in SEI capacity on lithiation vs. delithiation of carbon black. The relation $\tilde{\sigma}_{SEI,e} = \tilde{c}_{Li^+,SEI}^2$ approximates the effect of hopping conduction of electrons across percolating networks of inserted lithium ions in the SEI[74]. To visualize this phenomenon, the SEI layer can be thought of as a porous 3D isotropic lattice with some lattice sites (or defects) with the ability to accommodate absorbed lithium ions from the electrolyte. We use the term lattice loosely here since SEI contains amorphous phases.

Figure 6 presents the time profiles of current (6a), electronic conductivity of the SEI (6b), SEI thickness (6c), and lithium-ion concentration in the SEI (7d) for a simulation incorporating percolated hopping conduction ($\nu = 2$) for a nominal rate of C/10. During lithiation, the concentration of lithium ions increases in the SEI (6d), which randomly occupy a fraction of the lattice sites. Electrons hop across the layer to reach the outer surface of the SEI, at which point they react to form SEI. As the inserted ion (and mobile electron) concentration increases beyond a threshold value, the occupied sites form a 'spanning' cluster[53], which establishes long-range connectivity of the ions. This network allows the electrons to complete a hopping conduction event across the entire domain (illustrated as process 2 in Figure 1). According to standard percolation theory[75], the probability of obtaining long-range connectivity of hopping sites scales roughly as the square of the electron/ion concentration beyond a critical filling fraction of inserted ions.

                                    *Das et al.*

Upon reversal of the direction of the electric field in the SEI (i.e. reversal of the driving current), the outer film is depleted of lithium ions, and the concentration of ions drops instantaneously in the SEI as the effect of this depletion is felt throughout the layer (assuming fast transport of lithium ions in the SEI). This event causes the long-range connectivity of the percolating network of localized electronic states to be disrupted, leading to a large and instantaneous drop in electron conductivity (6b) per Equation (6) and breaking the symmetry of electron availability in the outer reaction film. This affects the exchange current and the overpotential of the solvent/electrolyte reduction reaction. The mixed-conducting SEI, acting like a diode, becomes highly passivating and shuts down the SEI current almost completely (6a) in the delithiation step, resulting in near-zero growth of the layer (6c).

Given the rapid change in SEI growth rate upon reversal of the sign of the current, we mention that an interfacial electron transfer effect, as proposed here, is a more plausible mechanism of directional asymmetry in SEI growth than a mechanism involving bulk transport by solid diffusion (i.e. within a carbon black particle). Although solid diffusion is also affected by the sign of current and can exhibit history dependence, solid diffusion is slow and cannot respond so quickly to current reversal.

The idea of a dependence of SEI conductivity on concentration and potential is not new. Many authors have considered ion diffusion through defects in the SEI to be the rate determining step for SEI growth by performing atomistic calculations[76–78] based on model compounds like $Li_2CO_3$ and LiF. However, experimental results such as the ones by Zhuang et al.[79] and Kobayashi et al.[80] suggested that growth kinetics are not just a function of ion diffusion but also electron conduction[81]. Shi et al.[76] speculated upon an electron leakage mechanism, which depends upon neutral lithium diffusion (i.e. ambipolar diffusion of $Li^+/e^-$ polarons) through defects in the SEI. In this study, we claim that the coupling between ionic concentration and electronic conduction across the SEI is what governs the post-first-cycle growth kinetics. To the best of the authors' knowledge, this paper presents the first experimentally validated electron conduction mechanism for post-first cycle SEI growth that is coupled to the inserted ion concentration within the layer.

As the SEI layer gets thicker with increasing cycle number, electron diffusion across the SEI layer may eventually become more difficult, as finite clusters of hopping sites no longer percolate across the layer. This picture for thicker SEI layers would be consistent with Peled's original theory of electron transport limited growth[82]. The SEI may also become porous with the growth reaction distributed over an internal surface area, influenced by the competing diffusion of electrons, ions, and solvent molecules in a heterogeneous composite[31-32].

***Interdependence of cycle time and driving current in SEI growth:*** Figure 7a describes the dependence of the second cycle SEI capacity during lithiation and delithiation as a function of the nominal C rate. The second-cycle SEI capacity decreases with C rate during lithiation and is independent of C rate during delithiation. This behavior has been primarily attributed to the higher cycle time spent for lower driving currents, allowing more time for SEI to grow[20]. However, if time was the only dominant factor, then the time-normalized SEI capacity should be independent of C rate, as the SEI reaction roughly goes through the same overpotential profile for different driving currents. However, Figure 7b depicts the average SEI current (i.e. the SEI capacity divided by the time per cycle) for lithiation increasing with total current with an average dimensionless slope of ~0.1. This theoretical result is consistent with experiment. This relationship signifies a nontrivial interdependence between cycle time and C rate in governing SEI growth kinetics.

We explore this interdependence further using parametric simulations for a range of values of electronic resistance ($R_{SEI,e}$), SEI kinetic rate constant ($k_{SEI}$) and intercalation charge-transfer coefficient ($\alpha_{int}$). The colormap plots in Figure S2 in Supplementary Material show the dependence of the average SEI growth rate on C rate for $v = 0$, 1, and 2. The $v = 0$ case (constant electronic conductivity) gives us the expected zero dependence on C rate; in other words, SEI growth is only dependent on time. For the $v = 1$ and $v = 2$ cases (ideal and non-ideal MIECs), the dependence is non-zero for most values in the parameter space. This result suggests that the model captures the time-current interdependence that persists for a range of current rates below C/5. For a fixed total current $I_{tot}$, an increase in the kinetic rate constant $k_{SEI}$ results in an increase in the SEI current $I_{SEI}$ and a decrease in the intercalation current $I_{int}$. A higher $I_{SEI}$ translates to a higher slope (a measure of time-current interdependence), which causes the color gradient to be primarily vertical and increasing with $k_{SEI}$. Since the average conductivity is higher for $v = 1$ than for $v = 2$, the time-current interdependence for the former is higher. We conclude that SEI capacity is a function of both the SEI current and the total cycle time, and the former increases with the nominal C rate.

## Conclusions

This modeling work is based on the experimental observations of Part I, which identifies strong dependencies of SEI growth to voltage, current magnitude, and current direction. In Part II, we develop a model of post-first-cycle SEI growth that is extensively validated across a range of low (C/100) to moderate (C/5) current rates. To explain the observed stark directional asymmetry in SEI growth between lithiation and delithiation, we develop a model that assumes a non-linear coupling of ion concentration and electron transfer. In this framework, the SEI behaves as a mixed ion-electron conductor, with the electron conductivity varying approximately as the square of lithium ion concentration in the SEI. This mechanism is consistent with hopping conduction of electrons across a percolating cluster of inserted lithium ions. We find that SEI growth is limited by electron availability in the outer reaction film. Delithiation of the electrode results in an instantaneous depletion of the lithium ion concentration in the SEI, breaking the symmetry of electron availability for SEI and leading to low growth on delithiation.

We have explored how the coupling between the concentration of lithium ions and electrons in the SEI can influence the electron conductivity, the exchange current, and overpotential of the electrolyte reduction reaction. From the results in Figures 4 and 5, $R_{SEI,e}$ is an important parameter determining the degree of directional asymmetry. Using the model, we find that $R_{SEI,e}$ influences the overpotential available for the electrolyte/solvent reduction reaction and acts to reduce the available overpotential as the resistance is increased. Most SEI models typically consider the resistance to be solely a function of the SEI thickness, but here we show that SEI thickness may be a much more complex function of transient variables such as the cell potential, electron concentration, current direction, current magnitude, and the ion absorption energy.

Our model has seven fitted parameters that influence the electrochemical kinetics of SEI growth. We have presented one set of parameter values in Table 1 that leads to the reproduction of the experimental outcome through coupling of the intercalation and electrolyte reduction reactions. Modifications to this framework include specifying electrolyte reduction/intercalation charge transfer coefficients that deviate from 0.5, changing the plane of growth (which would change the effect of concentration variables on the electrolyte/solvent reduction), assuming a different rate limiting step, changing the magnitude of the ion absorption energy, and assuming non-integer values of the scaling exponent $\nu$. However, considerable covariance exists between the parameters, making distinguishing between sets of parameter values difficult.

These results warrant a fresh look at the nature of the SEI layer in early-stage (i.e. post-first-cycle) capacity fade modeling efforts, accounting for the voltage and current dependencies within the span of a cycle. In this time regime, our results suggest a crucial coupling between ion and election transfer through SEI. This interaction also affects the interplay of cycle time and C rate in determining the overall average SEI capacity under different conditions, and model results predict the shift in the trend of C rate dependence of SEI capacity as the driving current is increased beyond C/5. This observation may have interesting implications for lithium-ion batteries operated at high charging rates, which is a major focus of research efforts in the field. Lastly, this model provides important insights into the fundamental nature of electrochemical SEI and could be incorporated into the degradation modules of battery modeling software such as MPET[83].

**Acknowledgements**

We thank Prof. Richard D. Braatz and Dr. Stephen J. Harris for insightful discussions. This work was supported by the Toyota Research Institute through D3BATT: Center for Data-Driven Design of Li-Ion Batteries (S. D.) and by the Thomas V. Jones Stanford Graduate Fellowship, and the National Science Foundation Graduate Research Fellowship under Grant No. DGE-114747 (P. M. A.).

*Das et al.*

**Table A1: Appendix**

| Symbol | Description | Unit |
|---|---|---|
| $dQ/dV$ | Differential capacity during cycling | mAh g$^{-1}$ V$^{-1}$ |
| $\Delta dQ/dV$ | Differential capacity due to isolated SEI growth | mAh g$^{-1}$ V$^{-1}$ |
| $\Delta\tilde{\phi}_{int}$ | Dimensionless potential drop (scaled to thermal voltage kT/e) at carbon/SEI interface due to (de)intercalation | - |
| $\Delta\tilde{\phi}_{int}^0$ | Intercalation open circuit potential, dependent on SoC | - |
| $\Delta\tilde{\phi}_{SEI}$ | Dimensionless potential drop driving electronic conduction across SEI; $\tilde{\phi}_{SEI}(x=\tilde{L}) - \tilde{\phi}_{SEI}(x=0)$ | - |
| $\Delta\tilde{\phi}_s$ | Dimensionless potential drop across the compact Stern layer at the SEI/electrolyte interface; $\tilde{\phi}_{elec} - \tilde{\phi}_{SEI}(x=\tilde{L})$ | - |
| $\tilde{\eta}_1$ | SEI growth / Electrolyte reduction overpotential; $\Delta\tilde{\phi}_s - \Delta\tilde{\phi}_{s,eq}$ | - |
| $\tilde{\eta}_2$ | Intercalation overpotential; $\Delta\tilde{\phi}_{int} - \Delta\tilde{\phi}_{int,eq}$ | - |
| $I_{int}$ | Current density per electrode area of (de)intercalation | A m$^{-2}$ |
| $I_{SEI}$ | Areal current of SEI growth | A m$^{-2}$ |
| $I_{tot}$ | Total areal current | A m$^{-2}$ |
| $\tilde{c}_{Li^+,SEI}$ | Dimensionless lithium ion concentration in SEI | - |
| $\tilde{c}_{e^-,SEI}$ | Dimensionless electron concentration in SEI | - |
| $\tilde{c}$ | Intercalated lithium filling fraction, measure of SoC | - |
| $R_{SEI,e}$ | Electronic resistance of SEI | $\Omega$ |
| $\tilde{L}$ | Dimensionless SEI thickness | - |

| $\widetilde{\sigma}_{SEI,e}$ | Dimensionless electronic conductivity in SEI | S m$^{-1}$ |
| $\nu$ | Scaling of e$^-$ conductivity to Li$^+$ concentration in SEI | - |

# References

(1) Hayashi, K.; Nemoto, Y.; Tobishima, S. I.; Yamaki, J. I. Mixed Solvent Electrolyte for High Voltage Lithium Metal Secondary Cells. *Electrochim. Acta* **1999**, *44* (14), 2337–2344.

(2) Joho, F.; Novák, P.; Spahr, M. E. Safety Aspects of Graphite Negative Electrode Materials for Lithium-Ion Batteries. *J. Electrochem. Soc.* **2002**, *149* (8), A1020.

(3) Zhang, Z.; Fouchard, D.; Rea, J. R. Differential Scanning Calorimetry Material Studies: Implications for the Safety of Lithium-Ion Cells. *J. Power Sources* **1998**, *70* (1), 16–20.

(4) Dahn, J. R.; Fuller, E. W.; Obrovac, M.; Vonsacken, U. Thermal-Stability of LixCoO2, LixNiO2 and Lambda-MnO2 and Consequences for the Safety of Li-Ion Cells. *Solid State Ionics* **1994**, *69* (3–4), 265–270.

(5) Lux, S. F.; Chevalier, J.; Lucas, I. T.; Kostecki, R. HF Formation in LiPF6-Based Organic Carbonate Electrolytes. *ECS Electrochem. Lett.* **2013**, *2* (12), A121–A123.

(6) Winter, M.; Besenhard, J. O. Electrochemical Intercalation of Lithium into Carbonaceous Materials. In *Lithium Ion Batteries : Fundamentals and Perhrmance*; 1998; pp 127–156.

(7) Aurbach, D.; Markovsky, B.; Weissman, I.; Levi, E.; Ein-Eli, Y. On the Correlation between Surface Chemistry and Performance of Graphite Negative Electrodes for Li Ion Batteries. *Electrochim. Acta* **1999**, *45* (1), 67–86.

(8) Nie, M.; Abraham, D. P.; Chen, Y.; Bose, A.; Lucht, B. L. Silicon Solid Electrolyte Interphase (SEI) of Lithium Ion Battery Characterized by Microscopy and Spectroscopy. *J. Phys. Chem. C* **2013**, *117* (26), 13403–13412.

(9) Nie, M.; Chalasani, D.; Abraham, D. P.; Chen, Y.; Bose, A.; Lucht, B. L. Lithium Ion Battery Graphite Solid Electrolyte Interphase Revealed by Microscopy and Spectroscopy. *J. Phys. Chem. C* **2013**, *117* (3), 1257–1267.

(10) Michan, A. L.; Leskes, M.; Grey, C. P. Voltage Dependent Solid Electrolyte Interphase Formation in Silicon Electrodes: Monitoring the Formation of Organic Decomposition Products. *Chem. Mater.* **2016**, *28* (1), 385–398.

(11) Xiao, A.; Yang, L.; Lucht, B. L.; Kang, S.-H.; Abraham, D. P. Examining the Solid Electrolyte Interphase on Binder-Free Graphite Electrodes. *J. Electrochem. Soc.* **2009**, *156* (4), A318.

(12) Nie, M.; Abraham, D. P.; Seo, D. M.; Chen, Y.; Bose, A.; Lucht, B. L. Role of Solution Structure in Solid Electrolyte Interphase Formation on Graphite with LiPF6 in Propylene Carbonate. *J. Phys. Chem. C* **2013**, *117* (48), 25381–25389.

(13)  Smith, A. J.; Burns, J. C.; Zhao, X.; Xiong, D.; Dahn, J. R. A High Precision Coulometry Study of the SEI Growth in Li/Graphite Cells. *J. Electrochem. Soc.* **2011**, *158* (9), A447–A452.

(14)  Smith, A. J.; Dahn, J. R. Delta Differential Capacity Analysis. *J. Electrochem. Soc.* **2012**, *159* (3), A290.

(15)  Bloom, I.; Cole, B. W.; Sohn, J. J.; Jones, S. A.; Polzin, E. G.; Battaglia, V. S.; Henriksen, G. L.; Motloch, C.; Richardson, R.; Unkelhaeuser, T.; et al. An Accelerated Calendar and Cycle Life Study of Li-Ion Cells. *J. Power Sources* **2001**, *101* (2), 238–247.

(16)  Peled, E. The Electrochemical Behavior of Alkali and Alkaline Earth Metals in Nonaqueous Battery Systems—The Solid Electrolyte Interphase Model. *J. Electrochem. Soc.* **1979**, *126* (12), 2047.

(17)  Broussely, M.; Herreyre, S.; Biensan, P.; Kasztejna, P.; Nechev, K.; Staniewicz, R. J. Aging Mechanism in Li Ion Cells and Calendar Life Predictions. *J. Power Sources* **2001**, *97–98*, 13–21.

(18)  Ploehn, H. J.; Ramadass, P.; White, R. E. Solvent Diffusion Model for Aging of Lithium-Ion Battery Cells. *J. Electrochem. Soc.* **2004**, *151* (3), A456.

(19)  Christensen, J.; Newman, J. A Mathematical Model for the Lithium-Ion Negative Electrode Solid Electrolyte Interphase. *J. Electrochem. Soc.* **2004**, *151* (11), A1977.

(20)  Pinson, M. B.; Bazant, M. Z. Theory of SEI Formation in Rechargeable Batteries: Capacity Fade, Accelerated Aging and Lifetime Prediction. *J. Electrochem. Soc.* **2013**, *160* (2), A243–A250.

(21)  Peled, E.; Menkin, S. Review—SEI: Past, Present and Future. *J. Electrochem. Soc.* **2017**, *164* (7), A1703–A1719.

(22)  Edstrom, K.; Herstedt, M.; Abraham, D. P. A New Look at the Solid Electrolyte Interphase on Graphite Anodes in Li-Ion Batteries. *J. Power Sources* **2006**, *153* (2), 380–384.

(23)  Broussely, M.; Biensan, P.; Bonhomme, F.; Blanchard, P.; Herreyre, S.; Nechev, K.; Staniewicz, R. J. Main Aging Mechanisms in Li Ion Batteries. *J. Power Sources* **2005**, *146*, 90–96.

(24)  Vetter, J.; Novák, P.; Wagner, M. R.; Veit, C.; Möller, K. C.; Besenhard, J. O.; Winter, M.; Wohlfahrt-Mehrens, M.; Vogler, C.; Hammouche, A. Ageing Mechanisms in Lithium-Ion Batteries. *J. Power Sources* **2005**, *147* (1–2), 269–281.

(25)  Verma, P.; Maire, P.; Novák, P. A Review of the Features and Analyses of the Solid Electrolyte Interphase in Li-Ion Batteries. *Electrochim. Acta* **2010**, *55* (22), 6332–6341.

(26)  Zhang, H. L.; Li, F.; Liu, C.; Tan, J.; Cheng, H. M. New Insight into the Solid Electrolyte Interphase with Use of a Focused Ion Beam. *J. Phys. Chem. B* **2005**, *109* (47), 22205–22211.

(27)  Lu, P.; Li, C.; Schneider, E. W.; Harris, S. J. Chemistry, Impedance, and Morphology

Evolution in Solid Electrolyte Interphase Films during Formation in Lithium Ion Batteries. *J. Phys. Chem. C* **2014**, *118* (2), 896–903.

(28)  Roder, F.; Braatz, R. D.; Krewer, U. Multi-Scale Simulation of Heterogeneous Surface Film Growth Mechanisms in Lithium-Ion Batteries. *J. Electrochem. Soc.* **2017**, *164* (11), E3335–E3344.

(29)  Burghaus, U.; Stephan, J.; Vattuone, L.; Rogowska, J. M. *A Practical Guide to Kinetic Monte Carlo Simulations and Classical Molecular Dynamics Simulations: An Example Book*; 2006.

(30)  Single, F.; Horstmann, B.; Latz, A. Revealing SEI Morphology: In-Depth Analysis of a Modeling Approach. *J. Electrochem. Soc.* **2017**, *164* (11), E3132–E3145.

(31)  Single, F.; Latz, A.; Horstmann, B. Identifying the Mechanism of Continued SEI Growth. *ChemSusChem* **2018**.

(32)  Shi, S.; Lu, P.; Liu, Z.; Qi, Y.; Hector, L. G.; Li, H.; Harris, S. J. Direct Calculation of Li-Ion Transport in the Solid Electrolyte Interphase. *J. Am. Chem. Soc.* **2012**, *134* (37), 15476–15487.

(33)  Arora, P.; White, R. E.; Doyle, M. Capacity Fade Mechanisms and Side Reactions in Lithium-Ion Batteries. *J. Electrochem. Soc.* **1998**, *145* (10), 3647.

(34)  See, K. A.; Lumley, M. A.; Stucky, G. D.; Grey, C. P.; Seshadri, R. Reversible Capacity of Conductive Carbon Additives at Low Potentials: Caveats for Testing Alternative Anode Materials for Li-Ion Batteries. *J. Electrochem. Soc.* **2017**, *164* (2), A327–A333.

(35)  Chan, C. K.; Ruffo, R.; Hong, S. S.; Cui, Y. Surface Chemistry and Morphology of the Solid Electrolyte Interphase on Silicon Nanowire Lithium-Ion Battery Anodes. *J. Power Sources* **2009**, *189*, 1132–1140.

(36)  Tasaki, K.; Goldberg, A.; Lian, J.-J.; Walker, M.; Timmons, A.; Harris, S. J. Solubility of Lithium Salts Formed on the Lithium-Ion Battery Negative Electrode Surface in Organic Solvents. *J. Electrochem. Soc.* **2009**, *156*, A1019.

(37)  Pan, J.; Cheng, Y. T.; Qi, Y. General Method to Predict Voltage-Dependent Ionic Conduction in a Solid Electrolyte Coating on Electrodes. *Phys. Rev. B - Condens. Matter Mater. Phys.* **2015**, *91* (13).

(38)  Benitez, L.; Seminario, J. M. Ion Diffusivity through the Solid Electrolyte Interphase in Lithium-Ion Batteries. *J. Electrochem. Soc.* **2017**, *164* (11), E3159–E3170.

(39)  Bard, A. J.; Faulkner, L. R.; York, N.; @bullet, C.; Brisbane, W.; Toronto, S. E. *ELECTROCHEMICAL METHODS Fundamentals and Applications*; 1944.

(40)  Biesheuvel, P. M.; Franco, A. A.; Bazant, M. Z. Diffuse Charge Effects in Fuel Cell Membranes. *J. Electrochem. Soc.* **2009**, *156* (2), B225.

(41)  Bazant, M. Z.; Chu, K. T.; Bayly, B. J. Current-Voltage Relations for Electrochemical Thin Films. **2005**, *65* (5), 1463–1484.

(42)  Newman, J.; Thomas-Alyea, K. *Electrochemical Systems*, Third Edit.; 2004; Vol. 7.

(43) Smith, R. B.; Bazant, M. Z. Multiphase Porous Electrode Theory. *J. Electrochem. Soc.* **2017**, *164* (11), E3291–E3310.

(44) Bazant, M. Z. Theory of Chemical Kinetics and Charge Transfer Based on Nonequilibrium Thermodynamics. *Acc. Chem. Res.* **2013**, *46* (5), 1144–1160.

(45) Biesheuvel, P. M.; van Soestbergen, M.; Bazant, M. Z. Imposed Currents in Galvanic Cells. *Electrochim. Acta* **2009**, *54* (21), 4857–4871.

(46) Biesheuvel, P. M.; Fu, Y.; Bazant, M. Z. Diffuse Charge and Faradaic Reactions in Porous Electrodes. *Phys. Rev. E - Stat. Nonlinear, Soft Matter Phys.* **2011**, *83* (6).

(47) Andersen, M. B.; Van Soestbergen, M.; Mani, A.; Bruus, H.; Biesheuvel, P. M.; Bazant, M. Z. Current-Induced Membrane Discharge. *Phys. Rev. Lett.* **2012**, *109* (10), 1–5.

(48) Moran, J. H.; Wheat, P. M.; Posner, J. D. Locomotion of Electrocatalytic Nanomotors Due to Reaction Induced Charge Autoelectrophoresis. *Phys. Rev. E* **2010**, *81*, 065302.

(49) Riess, I. Mixed Ionic–electronic Conductors—material Properties and Applications. *Solid State Ionics* **2003**, *157* (1–4), 1–17.

(50) Bergman, D. J.; Stroud, D. Physical Properties of Macroscopically Inhomogeneous Media. *Solid State Phys. - Adv. Res. Appl.* **1992**, *46* (C), 147–269.

(51) Stauffer, D.; Aharony, a. Introduction to Percolation Theory. *Computer.* 1994, p 192.

(52) Torquato, S. *Random Heterogeneous Materials: Microstructure and Macroscopic Properties*; 2012; Vol. XXXIII.

(53) Rintoul, M. D.; Torquato, S. Precise Determination of the Critical Threshold and Exponents in a Three-Dimensional Continuum Percolation Model. *J. Phys. A. Math. Gen.* **1997**, *30* (16).

(54) Clerc, J. P.; Giraud, G.; Laugier, J. M.; Luck, J. M. The Electrical Conductivity of Binary Disordered Systems, Percolation Clusters, Fractals and Related Models. *Adv. Phys.* **1990**, *39* (3), 191–309.

(55) Kozlov, B.; Lagus, M. Universality of 3D Percolation Exponents and First-Order Corrections to Scaling for Conductivity Exponents. *Phys. A Stat. Mech. its Appl.* **2010**, *389* (23), 5339–5346.

(56) Behrens, S. H.; Grier, D. G. The Charge of Glass and Silica Surfaces. *J. Chem. Phys.* **2001**, *115* (14), 6716–6721.

(57) Deng, D.; Dydek, E. V.; Han, J. H.; Schlumpberger, S.; Mani, A.; Zaltzman, B.; Bazant, M. Z. Overlimiting Current and Shock Electrodialysis in Porous Media. *Langmuir* **2013**, *29* (52), 16167–16177.

(58) Bazant, M. Z.; Kilic, M. S.; Storey, B. D.; Ajdari, A. Towards an Understanding of Induced-Charge Electrokinetics at Large Applied Voltages in Concentrated Solutions. *Adv. Colloid Interface Sci.* **2009**, *152* (1–2), 48–88.

(59) Lyklema, J.; Duval, J. F. L. Hetero-Interaction between Gouy-Stern Double Layers:

Charge and Potential Regulation. *Adv. Colloid Interface Sci.* **2005**, *114–115*, 27–45.

(60)   Duval, J. F. L.; Leermakers, F. A. M.; Van Leeuwen, H. P. Electrostatic Interactions between Double Layers: Influence of Surface Roughness, Regulation, and Chemical Heterogeneities. *Langmuir* **2004**, *20* (12), 5052–5065.

(61)   Bai, P.; Cogswell, D. A.; Bazant, M. Z. Suppression of Phase Separation in LiFePO4 Nanoparticles during Battery Discharge. *Nano Lett.* **2011**, *11* (11), 4890–4896.

(62)   Dahn, J. R.; Fong, R.; Spoon, M. J. Suppression of Staging in Lithium-Intercalated Carbon by Disorder in the Host. *Phys. Rev. B* **1990**, *42* (10), 6424–6432.

(63)   Lukatskaya, M. R.; Dunn, B.; Gogotsi, Y. Multidimensional Materials and Device Architectures for Future Hybrid Energy Storage. *Nat. Commun.* **2016**, *7*, 1–13.

(64)   Zeng, Y.; Smith, R. B.; Bai, P.; Bazant, M. Z. Simple Formula for Marcus-Hush-Chidsey Kinetics. *J. Electroanal. Chem.* **2014**, *735*, 77–83.

(65)   Henstridge, M. C.; Laborda, E.; Wang, Y.; Suwatchara, D.; Rees, N.; Molina, Á.; Martínez-Ortiz, F.; Compton, R. G. Giving Physical Insight into the Butler-Volmer Model of Electrode Kinetics: Application of Asymmetric Marcus-Hush Theory to the Study of the Electroreductions of 2-Methyl-2-Nitropropane, Cyclooctatetraene and Europium(III) on Mercury Microelectrodes. *J. Electroanal. Chem.* **2012**, *672*, 45–52.

(66)   Marcus, R. A. On the Theory of Oxidation-Reduction Reactions Involving Electron Transfer. I. *J. Chem. Phys.* **1956**, *24* (5), 966–978.

(67)   Henstridge, M. C.; Laborda, E.; Compton, R. G. Asymmetric Marcus-Hush Model of Electron Transfer Kinetics: Application to the Voltammetry of Surface-Bound Redox Systems. *J. Electroanal. Chem.* **2012**, *674*, 90–96.

(68)   Zeng, Y.; Bai, P.; Smith, R. B.; Bazant, M. Z. Simple Formula for Asymmetric Marcus-Hush Kinetics. *J. Electroanal. Chem.* **2015**, *748*, 52–57.

(69)   Newman, J.; Thomas, K. E.; Hafezi, H.; Wheeler, D. R. Modeling of Lithium-Ion Batteries. *J. Power Sources* **2003**, *119–121*, 838–843.

(70)   Botte, G. G.; Subramanian, V. R.; White, R. E. Mathematical Modeling of Secondary Lithium Batteries. *Electrochim. Acta* **2000**, *45* (15–16), 2595–2609.

(71)   Doyle, M.; Fuller, T.; Newman, J. Modeling of Galvanostatic Charge and Discharge of the Lithium/Polymer/Insertion Cell. *J. Electrochem. Soc.* **1993**, *140* (6), 1526–1533.

(72)   Chidsey, C. E. Free Energy and Temperature Dependence of Electron Transfer at the Metal-Electrolyte Interface. *Science* **1991**, *251* (4996), 919–922.

(73)   Mott, N. F. The Metal-Insulator Transition in Extrinsic Semiconductors. *Adv. Phys.* **1972**, *21* (94), 785–823.

(74)   Bazant, M. Z. Largest Cluster in Subcritical Percolation. *Phys. Rev. E - Stat. Physics, Plasmas, Fluids, Relat. Interdiscip. Top.* **2000**, *62* (2), 1660–1669.

(75)   Christensen, K. Percolation Theory. *Thesis* **2002**, 1–40.

(76)  Shi, S.; Qi, Y.; Li, H.; Hector, L. G. Defect Thermodynamics and Diffusion Mechanisms in Li $_2$ CO $_3$ and Implications for the Solid Electrolyte Interphase in Li-Ion Batteries. *J. Phys. Chem. C* **2013**, *117* (17), 8579–8593.

(77)  Li, Y.; Leung, K.; Qi, Y. Computational Exploration of the Li-Electrode | Electrolyte Interface in the Presence of a Nanometer Thick Solid-Electrolyte Interphase Layer Published as Part of the Accounts of Chemical Research Special Issue " Nanoelectrochemistry " . *Acc. Chem. Res.* **2016**, *49*, 2363–2370.

(78)  Mizusaki, J.; Tagawa, H.; Saito, K.; Uchida, K.; Tezuka, M. Lithium Carbonate as a Solid Electrolyte. *Solid State Ionics* **1992**, *53–56* (PART 2), 791–797.

(79)  Zhuang, G. V.; Chen, G.; Shim, J.; Song, X.; Ross, P. N.; Richardson, T. J. Li2CO3in LiNi0.8Co0.15Al0.05O2cathodes and Its Effects on Capacity and Power. *J. Power Sources* **2004**, *134* (2), 293–297.

(80)  Kobayashi, H.; Shikano, M.; Koike, S.; Sakaebe, H.; Tatsumi, K. Investigation of Positive Electrodes after Cycle Testing of High-Power Li-Ion Battery Cells. I. An Approach to the Power Fading Mechanism Using XANES. *J. Power Sources* **2007**, *174* (2), 380–386.

(81)  Winter, M. The Solid Electrolyte Interphase-The Most Important and the Least Understood Solid Electrolyte in Rechargeable Li Batteries. *Zeitschrift für Phys. Chemie* **2009**, *223* (10–11), 1395–1406.

(82)  Peled, E. The Electrochemical Behavior of Alkali and Alkaline Earth Metals in Nonaqueous Battery Systems—The Solid Electrolyte Interphase Model. *J. Electrochem. Soc.* **1979**, *126* (1), 2047.

(83)  Smith, R. B.; Bazant, M. Z. MPET : Multiphase Porous Electrode Theory Software. *J. Electrochem. Soc.* **2017**.

(84)  Nie, M.; Abraham, D. P.; Seo, D. M.; Chen, Y.; Bose, A.; Lucht, B. L. Role of Solution Structure in Solid Electrolyte Interphase Formation on Graphite with LiPF 6 in Propylene Carbonate. *J. Phys. Chem. C* **2013**, *117* (48), 25381–25389.

| Symbol | Description | Value/Unit | Determination |
|--------|-------------|------------|---------------|
| $k_{SEI}$ | Exchange current prefactor for SEI | $2.5 \times 10^{-9}$ A m$^{-2}$ | Fitted |
| $k_{int}$ | Exchange current prefactor for (de)intercalation | $1.1 \times 10^{-2}$ A m$^{-2}$ | Fitted |
| $\Delta\phi^0_{SEI}$ | Equilibrium redox potential of SEI | 0.73 V (vs. Li/Li$^+$) | Fitted |
| $R^0_{SEI,e}$ | Initial SEI electronic resistance | 11.2 $\Omega$ | Fitted |
| $R^0_{SEI,Li^+}$ | Initial SEI ionic resistance | 0.03 $\Omega$ | Fitted |
| $\beta$ | Thickness scaling for changing porosity | 1.21 | Fitted |
| $E_{ads}$ | Li$^+$ ion absorption energy in SEI | 3.6 eV | Fitted |
| A | Parameter in open circuit potential expression for carbon black in Eq. 9 | -0.17 | Fitted |
| B | Parameter in open circuit potential expression for carbon black in Eq. 9 | -0.42 | Fitted |
| C | Parameter in open circuit potential expression for carbon black in Eq. 9 | -0.48 | Fitted |
| $L_0$ | Initial post-first cycle SEI thickness | 30 nm[84] | Assumed |
| $\alpha_{int}$ | Butler Volmer charge transfer coefficient for (de)intercalation | 0.50 | Assumed |
| $\alpha_{SEI}$ | Butler Volmer charge transfer coefficient for SEI growth | 0.50 | Assumed |
| $A_{SEI}$ | Specific surface area of SEI (assumed equal to that of super P carbon black electrode) | ~62 m$^2$ g$^{-1}$ | Supplier spec |

**Table 1:** List of all model parameters

**Figure 1:** Schematic of 1D uniform SEI layer on a carbon negative electrode depicting dominant phenomena that can affect outer SEI growth kinetics. *Process 1*: (a) electron incorporation into SEI from carbon (b) electron conduction across SEI (c) electron reaction to form new SEI. *Process 2*: (a) Absorption and desolvation of lithium ion from electrolyte (b) diffusion of lithium ion across SEI (c) intercalation of lithium into carbon. *Process 3*: (a) Solvent incorporation in SEI (b) Solvent diffusion across SEI. Image not to scale, SEI thickness is exaggerated to highlight relevant physical processes.

**Figure 2**: Theoretical dependence of conductivity as a function of the fraction of occupied sites (analogous to bond concentration) for random bond-diluted networks (filled circles) and correlated addition to uniform spanning tree model (hollow circles) on the square lattice averaged over 25 realizations on the 50x50x50 lattice. This behavior can be approximated by assuming the conductivity to vary roughly as the square of the fraction of occupied sites (superimposed dashed line). Base figure reproduced from Chubynsky M. V., Thorpe M. F., Phys. Rev. E 71, 056105 (2005).

**Figure 3**: Characteristics of SEI with constant electronic conductivity under constant current C/10 (a) SEI and intercalation current distribution. (b) Evolution of dimensionless SEI thickness as a function of cycle time during lithiation and delithiation.

**Figure 4**: Parametric analysis of the degree of directional asymmetry of SEI growth for various models, SEI reaction rate constants, SEI resistances and intercalation charge-transfer coefficients

for percolation conduction model of SEI. The nominal C rate plotted here is C/50. The dotted regions include experimental range of values within 95% CI.

**Figure 5**: $\Delta dQ/dV$ comparisons of theoretical predictions with experiments between cycle 2 and the baseline cycle for five different nominal C rates: (a) C/100, (b) C/50, (c) C/20, (d) C/10, and (e) C/5. The orange and blue set of lines in each subplot represent delithiation and lithiation, respectively. Dotted lines are theoretical predictions, and cell-averaged experimental values are represented using solid lines.

**Figure 6**: Characteristics of percolation model of electron conduction in SEI layer under constant nominal C rate of C/10 (a) SEI and intercalation currents distribution (b) Dynamic behavior of SEI conductivity (c) Evolution of dimensionless SEI thickness (normalized to thickness after first cycle) (d) concentration of lithium ions in SEI during lithiation and delithiation (normalized to electrolyte ion concentration).

**Figure 7**: Comparison of theoretical predictions to experimental values of second-cycle SEI growth as a function of current and current direction (a) Dependence of second-cycle SEI growth on nominal C rate. These capacities are calculated from the integral of the $\Delta dQ/dV$ curves in Figure 6. (b) Dependence of second-cycle time-normalized 'average' SEI growth rate on nominal C rate. The average SEI growth rate is calculated by dividing SEI growth by the actual time per cycle. In both subplots, the error bars in experimental data represent 95% confidence intervals (CI) of the mean.

**Supplementary Material Figure Captions**

**Figure S1**: Potential dependence of $\Delta$dQ/dV compared with experiments between cycle 2 and the baseline cycle for SEI exhibiting (a) non-ideal, percolating MIEC  (b) ideal MIEC, and (c) constant electronic conductivity, for a nominal total current C/10.

**Figure S2**: Parametric analysis of the slope of the average SEI growth rate when plotted against the nominal C rate, as predicted by the theory for a wide range of SEI resistances and models with different correlations of ion-electron conduction in SEI. White dotted regions include experimental values within 95% CI.

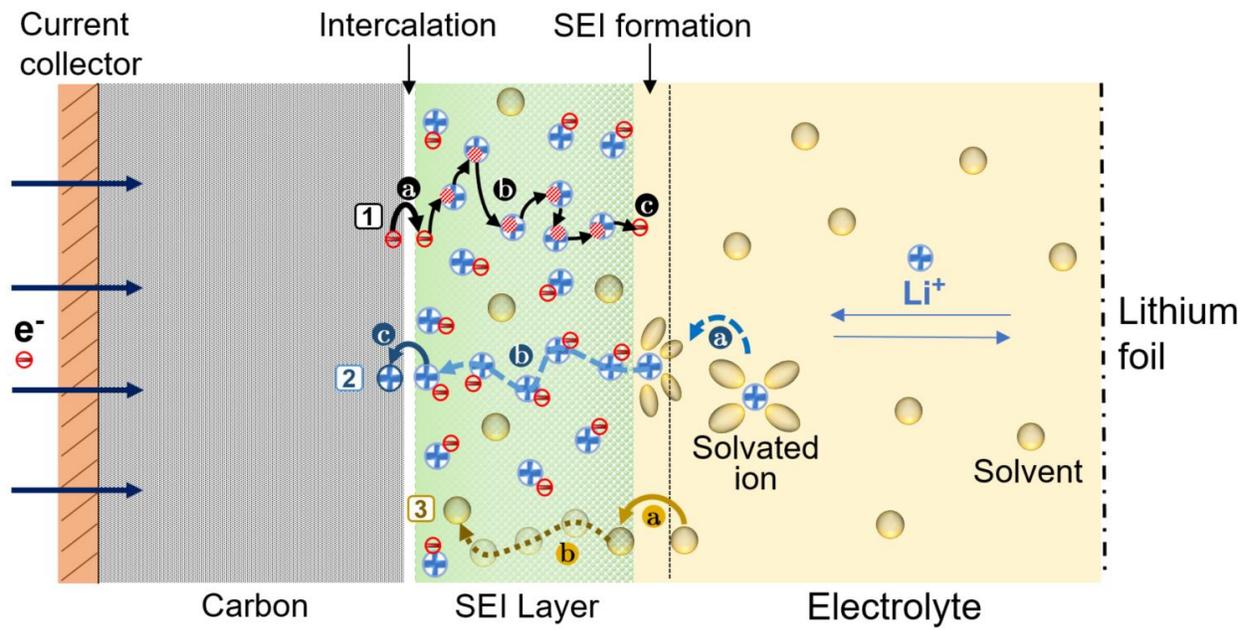

Current collector

Intercalation

SEI formation

e⁻

Li⁺

Lithium foil

Solvated ion

Solvent

Carbon

SEI Layer

Electrolyte

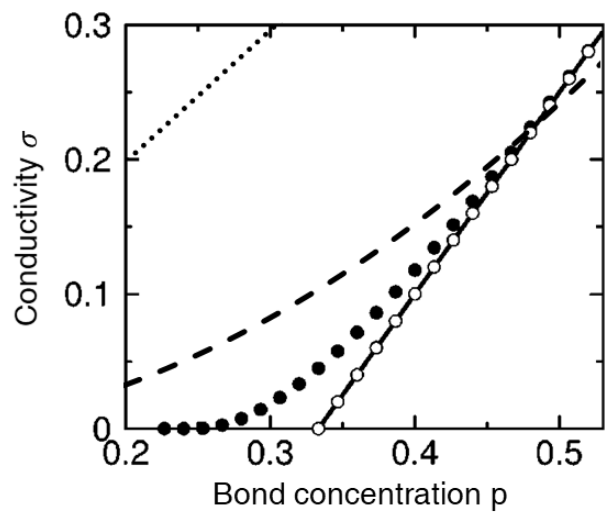

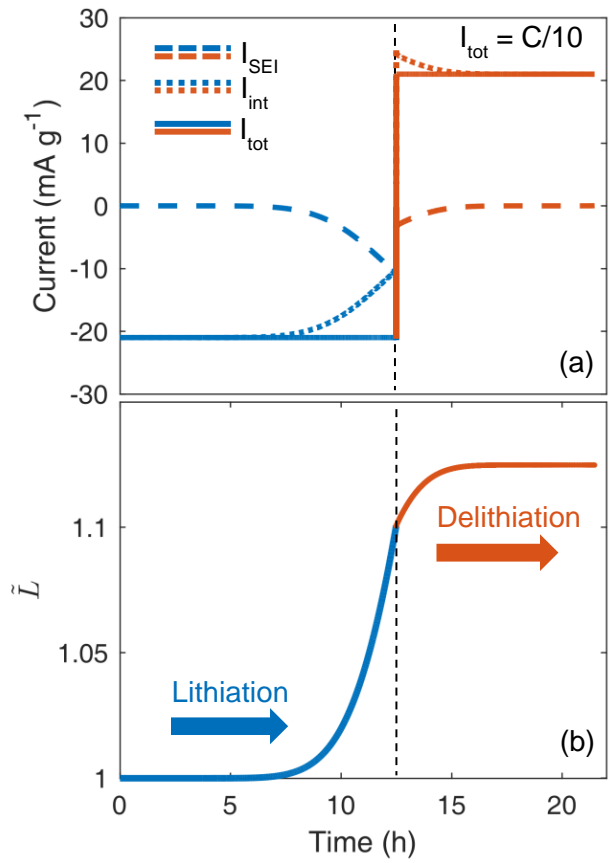

                                                        *Das et al.*

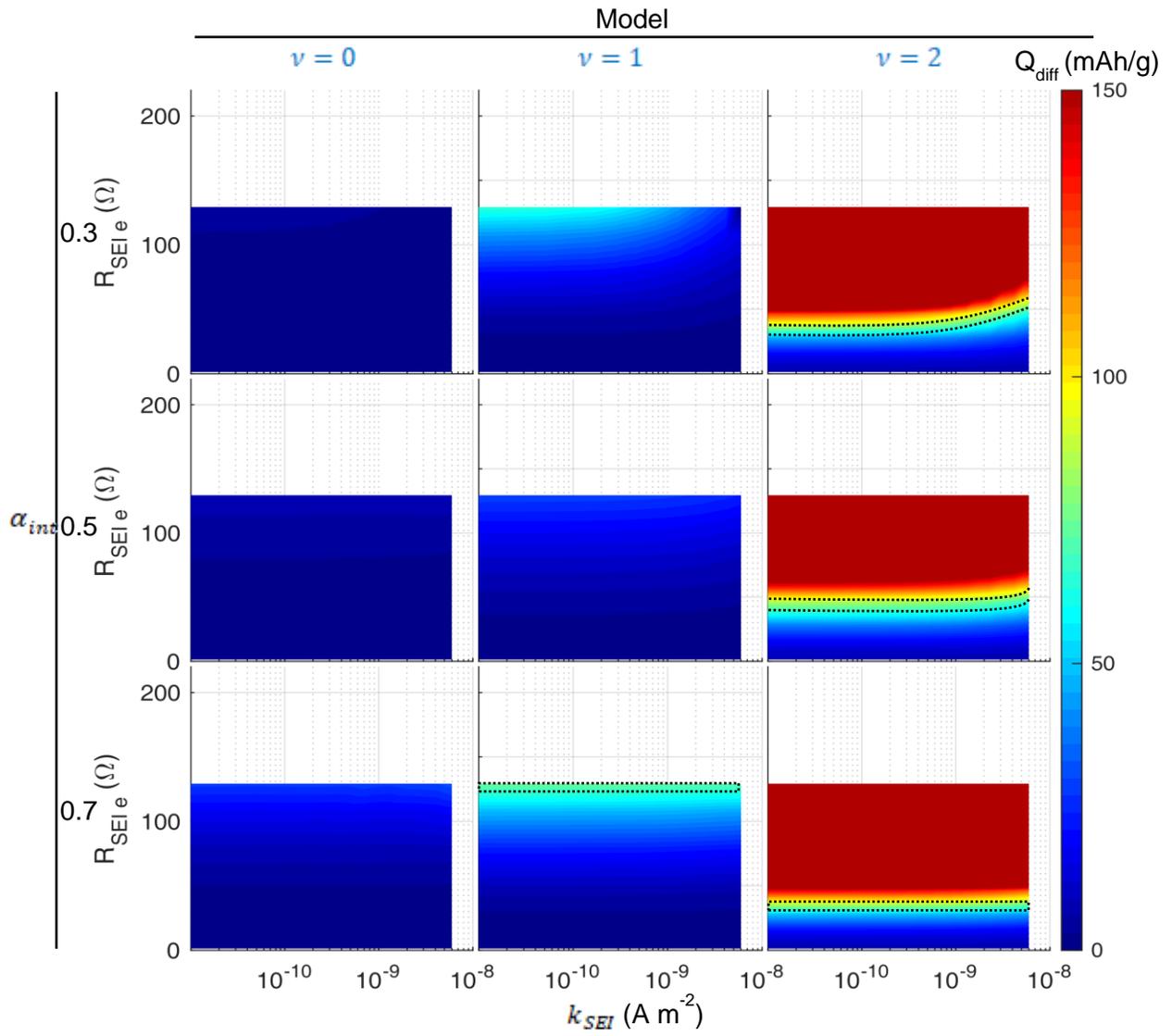

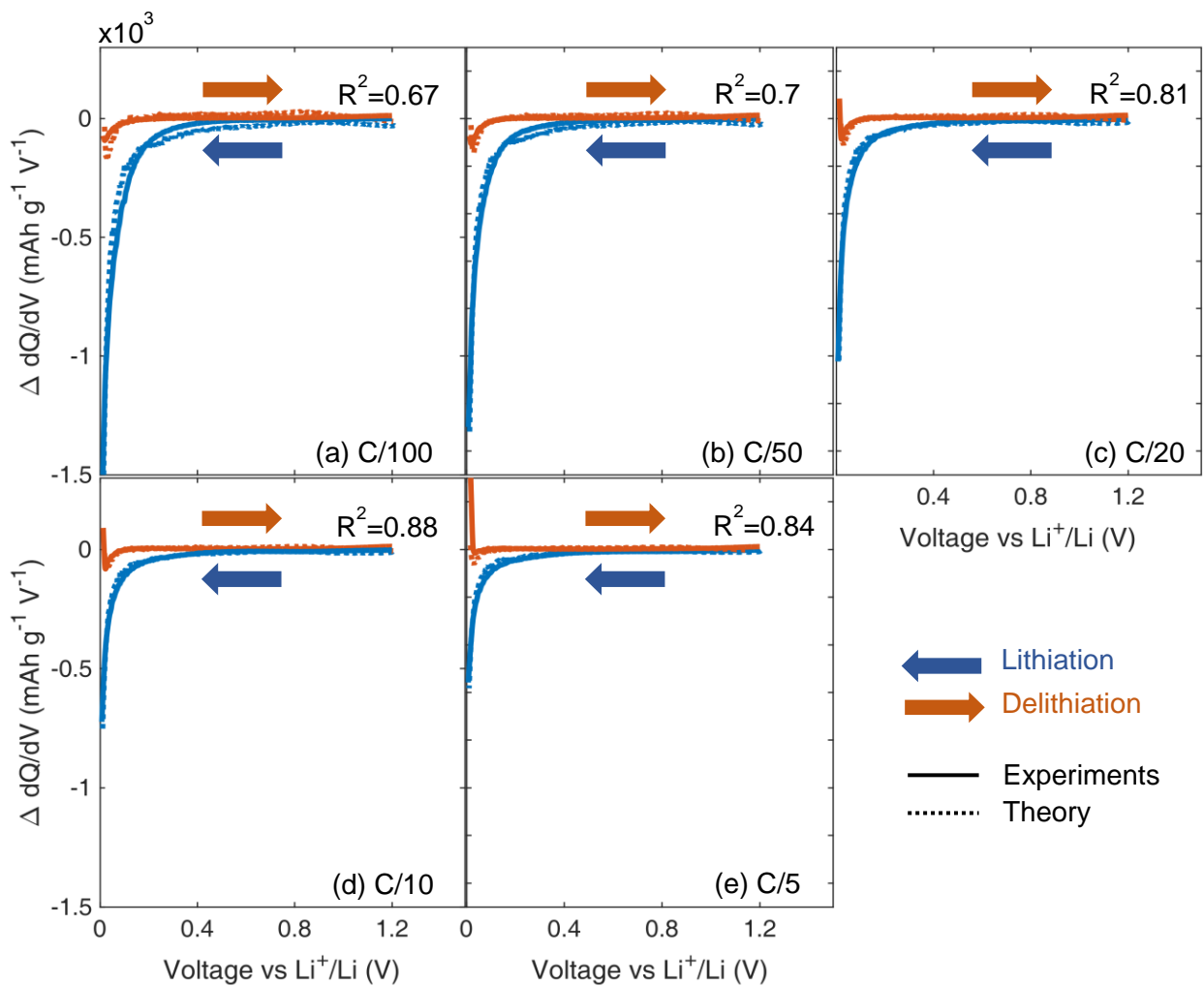

                                                    *Das et al.*

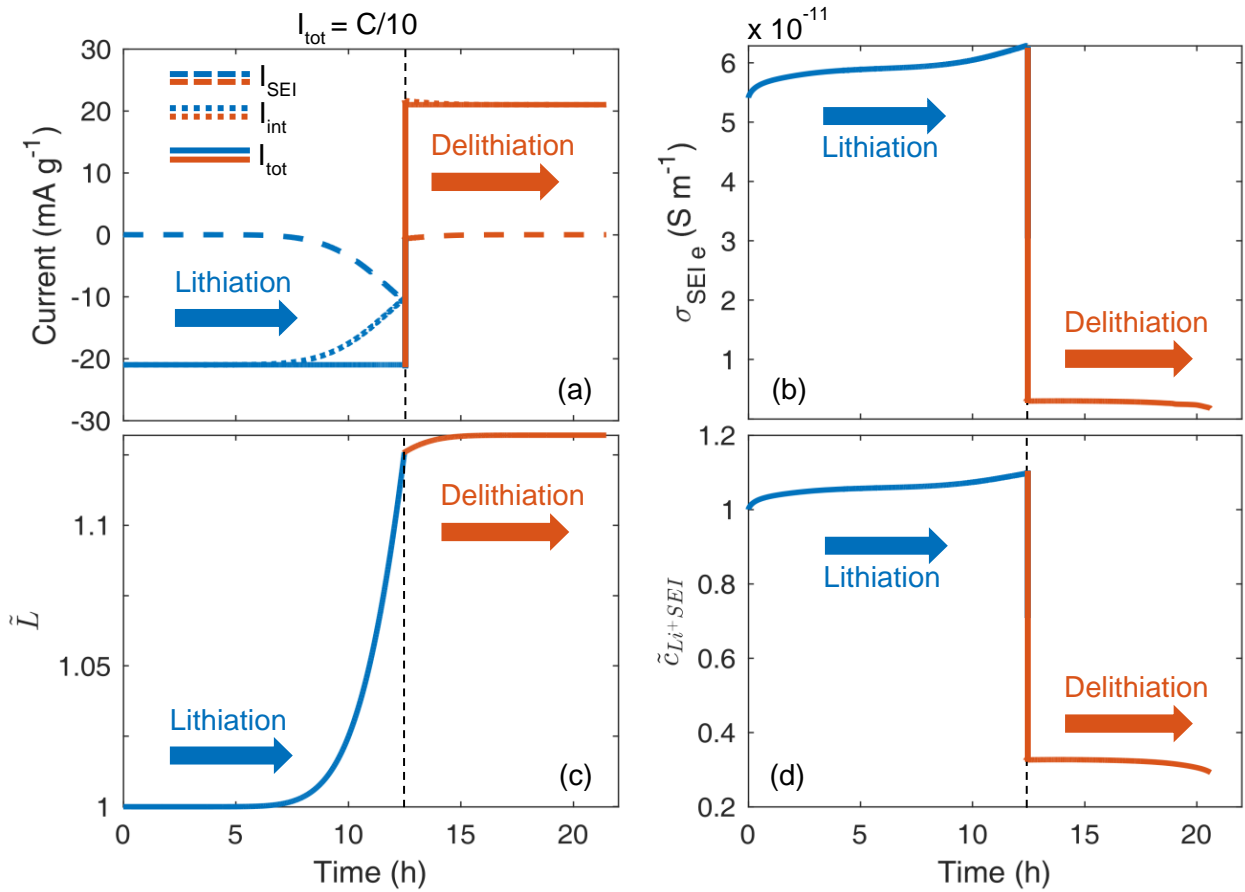

 

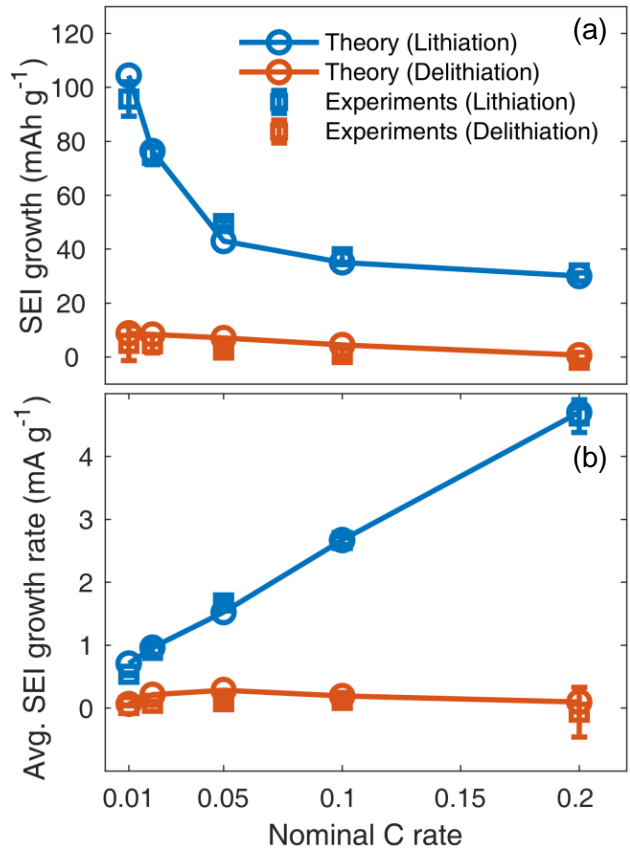

 *Das et al.*

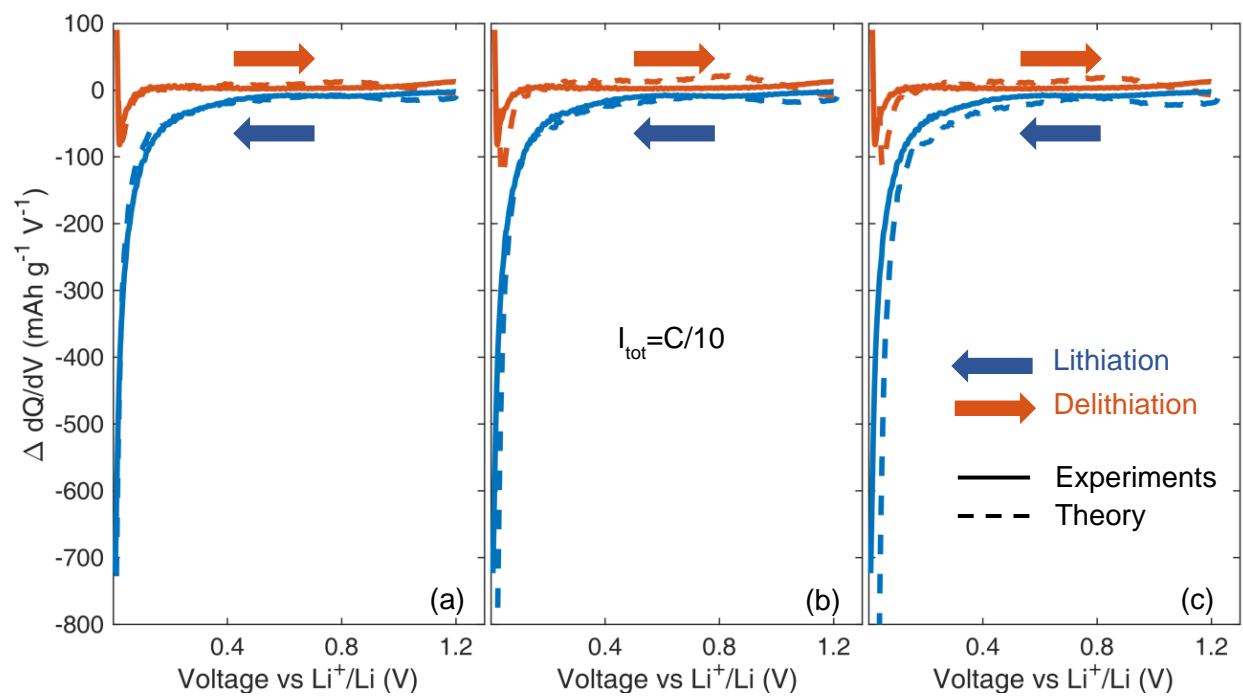

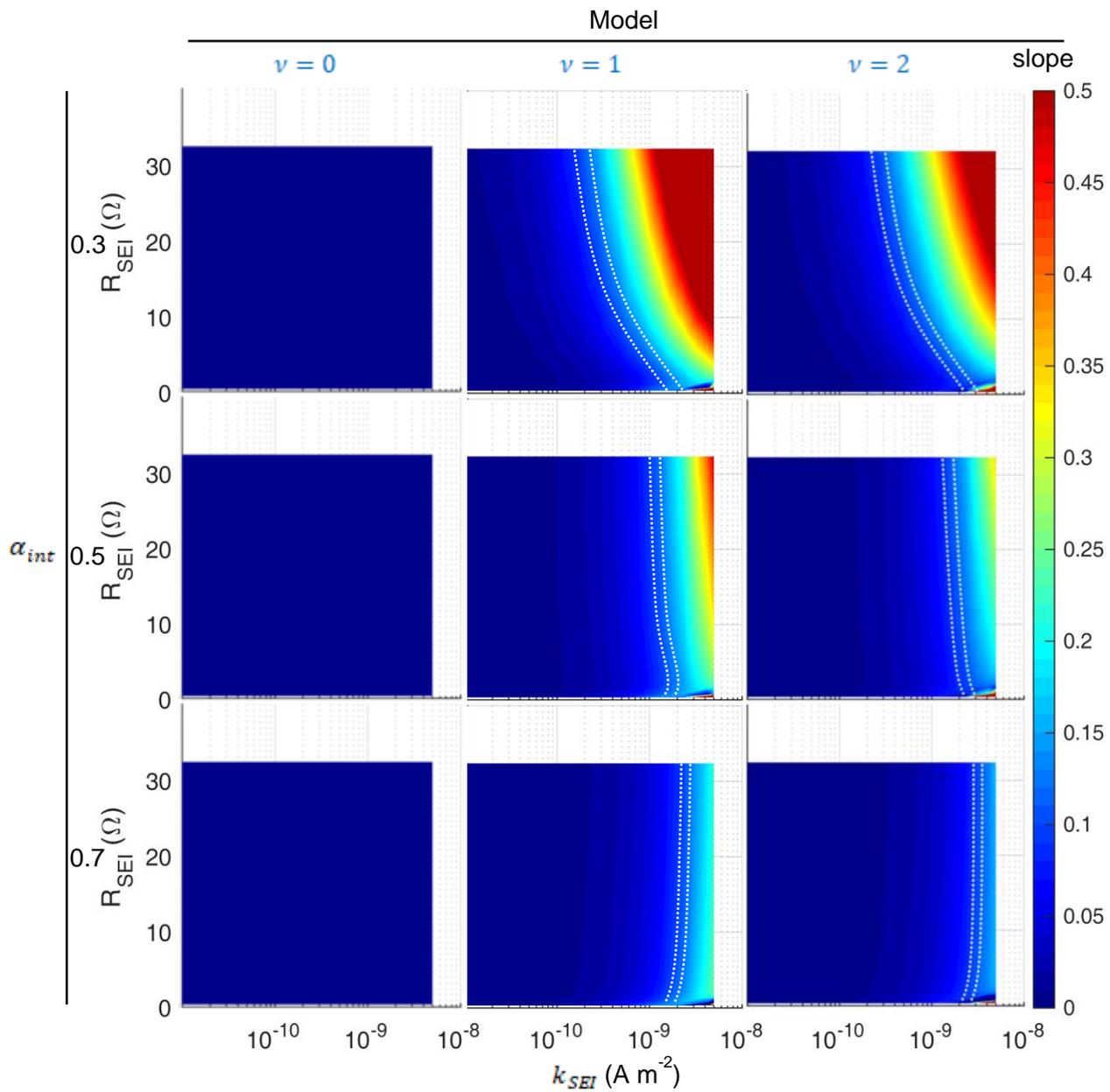